# Indexing magnetic structures and crystallographic distortions from powder diffraction: Brillouin zone indexing

## A.S. Wills[1,2]

[1]UCL, Department of Chemistry, 20 Gordon Street, London, UK, WC1H 0AJ
[2]London Centre for Nanotechnology, 17-19 Gordon Street, London, WC1H 0AH

[*] Contact author; e-mail: a.s.wills@ucl.ac.uk

**Abstract.** A considerable challenge is faced by researchers wishing to identify the propagation vector(s) associated with a magnetic structure or a lattice distortion from powder diffraction data, due to the severe destruction of information by powder averaging and search algorithms based on extracted peak positions. In this article a new method is introduced that is based on the points, lines and planes of the Brillouin zone of the crystal structure before the transition. These correspond to different classifications of the translational symmetry of the resultant order and of the Bloch wave that is used to describe the magnetic structure or phonon mode.
An automated search algorithm for the study of magnetic structures is described based on reverse-Monte Carlo refinement of moment orientations for a given k vector that markedly reduces the number of false-positives and allows the straight forward analysis of systems that contain contributions from several unrelated k vectors.

## 1. Introduction

The determination of the propagation vector(s) associated with a magnetic structure or a lattice distortion from powder diffraction data is a major challenge in powder diffraction, as much information is destroyed by powder averaging. An additional difficulty arises from the way that procedures to index the relevant diffraction peaks are decoupled from the nature of the physical processes that drives the phase transition: indexing is often carried out as a mathematical problem based on extracted peak positions and the calculation of the positions of reflections for simple trial cells. Other methods based on the formalism of a propagation vector, k vector, enable the exploration of both commensurate and incommensurate trial structures, either following grid searches or non-linear procedures but have difficulties with systems that involve several k vectors or poor data.[1]

Despite the high symmetry points of the Brillouin zone (BZ) being one of the basic concepts in solid state physics[2], they have not been systematically applied before to the indexing of magnetic structures and the propagation vectors associated with a displacive crystallographic phase transition. Their applicability arises from the restriction of magnetic structures and phonons to following the equation of a plane wave in accordance with Bloch's theorem. The k vector then defines the different classes of the translational symmetry of the resultant structure and in many cases the observed propagation vectors correspond to the different symmetry points, lines and planes in the Brillouin zone of the crystal structure before the ordering transition.

In this article we demonstrate how the Brillouin zone can be used to construct trial k vectors, and introduce a procedure for the determination of the k vector of a magnetic powder diffraction spectrum based on the reverse Monte-Carlo refinement of suitably located magnetic moments. Together these techniques enable the characterisation of complex systems, such as those with several propagation vectors which are not symmetry related, that may otherwise have been incorrectly characterised by a general point in the Brillouin zone.

## 2. The reciprocal lattice

Consider a lattice in direct space defined by the primitive translations $\mathbf{a}_1, \mathbf{a}_2, \mathbf{a}_3$. The position vector of any lattice point of that Bravais lattice is given by

$$\mathbf{T}_{n_1 n_2 n_3} = n_1 \mathbf{a}_1 + n_2 \mathbf{a}_2 + n_3 \mathbf{a}_3, \tag{1}$$

where $n_1$-$n_3$ are integers. The associated reciprocal lattice vectors are given by

$$\mathbf{b}_1 = 2\pi \frac{\mathbf{a}_2 \times \mathbf{a}_3}{\mathbf{a}_1 \cdot (\mathbf{a}_2 \times \mathbf{a}_3)} \tag{2}$$

$$\mathbf{b}_2 = 2\pi \frac{\mathbf{a}_3 \times \mathbf{a}_1}{\mathbf{a}_1 \cdot (\mathbf{a}_2 \times \mathbf{a}_3)} \tag{3}$$

$$\mathbf{b}_3 = 2\pi \frac{\mathbf{a}_1 \times \mathbf{a}_2}{\mathbf{a}_1 \cdot (\mathbf{a}_2 \times \mathbf{a}_3)} \tag{4}$$

and are the basic vectors of the Bravais lattice in reciprocal space, or **k** space. The reciprocal lattice may then be defined as the set of wave vectors **k** that satisfies the equation

$$e^{i\mathbf{kT}} = 1, \tag{5}$$

where **T** is a real space lattice vector defined in equation (1).

## 2. The Brillouin zone

The concept of the Brillouin zones was developed from Bloch's theory. Ignoring the correlations between electrons, this is equivalent to the *'independent particle approximation'* of Hartree-Fock, where every electron has a separate wave function that satisfies the Schrödinger equation:

$$-\frac{\hbar^2}{2m}\left(\frac{\partial^2}{\partial x^2}+\frac{\partial^2}{\partial y^2}+\frac{\partial^2}{\partial z^2}\right)\psi + V\psi = E\psi, \qquad (6)$$

where the potential $V$ has the symmetry of the lattice. Whereas in atoms and molecules the Eigenvalues of (1) are well separated, in crystalline solids they form a continuous manifold and must be characterised by continuously varying parameters as well as discrete points. [2]

The allowed solutions of (6) have the form of a Bloch wave:

$$\Psi_{\mathbf{k}}(\mathbf{r}) = u_{\mathbf{k}}(\mathbf{r})\exp(i\mathbf{kr}), \qquad (7)$$

where $\mathbf{k}$ is a reciprocal vector that may be defined using the basis vectors $\mathbf{a}_1$-$\mathbf{a}_3$ of the Bravais lattice. $u_{\mathbf{k}}(\mathbf{r})$ is a periodic function with the same periodic properties as $V(\mathbf{r})$. The solution $\psi_{\mathbf{k}}(\mathbf{r})$ has the form of a plane wave that is modulated by the periodic function $u_{\mathbf{k}}(\mathbf{r})$. Moving away from electronic wavefunctions, *per se*, magnetic structures or phonons are also the Eigenfuncitons of periodic Hamiltonians and are characterised by a sum over different Bloch waves

$$\Psi_{\mathbf{k}}(\mathbf{r}_i) = \sum_{k,\alpha} \psi_{\mathbf{k}}^{\alpha}(\mathbf{r}_j)\exp(i\mathbf{kt}_{ij}), \qquad (8)$$

where $\mathbf{t}_{ij}=\mathbf{r}_j-\mathbf{r}_i$ , $\mathbf{k}$ is the wavevector, and $\alpha$ is the index of the Bloch wave. [3-5]

The first BZ then defines a set of unique k vectors that satisfy (7) and cannot be made shorter by adding translation vectors of the reciprocal lattice. The different possible k vectors of the BZ may then be characterised into different points, lines and planes based on their small representations of the group of the wave vector, and the symmetry elements that leave the k vector invariant. [2] If the wave vector lies in a general position, the group of the wave vector will contain only the identity operation. For other values, the group of k contains more than the identity, for example in a tetragonal crystal where the k vector lies along the four-fold axis.

## 3. Application of Brillouin zone indexing to magnetic powder diffraction

The method is based on making the refinement process as physical as possible in order to counter the loss of information from powder averaging. Introducing physics at even a basic level greatly reduces the parameter space to be searched and makes tractable the analysis of systems that cannot be achieved by other methods, and follows the philosophy that we developed of refining magnetic structures in terms of the mixing coefficients of basis vectors in SARA$h$[6]. When applied to magnetic structures there are two main features of this process: the generation of the trial k vectors following the symmetry points of the Brillouin zone, and their rigorous testing using moment vectors and a reverse-Monte Carlo-Rietveld algorithm.

The technique was originally developed for the analysis of the magnetic scattering in the frustrated magnets β-Mn[7] and gadolinium gallium garnet (GGG) [8], as the competition between the magnetic interactions make the k vector impossible to guess, without extremely difficult theoretical calculations. Typically, such calculations are not possible as the magnetic structure provides the necessary starting information.

### 3.1. Generation of k vectors

There are several conventions that are used to define the high symmetry points, lines and planes of the Brillouin zone. For simplicity we use that of Kovalev [9] as it allows direct translation of the observed ordering wavevector to the representational analysis calculations of the basis vectors of the different possible magnetic structures (a technique that we have applied to several systems, including $Gd_2Sn_2O_7$ [10] and $CuB_2O_4$ [11]). The search procedure is based on trying the high symmetry points, then the lines and then the planes. This search order prevents the true symmetry from being misunderstood when the high symmetry points are coincident with lower symmetry lines.

There are typically only a few commensurate symmetry points and these are searched sequentially. If the diffraction spectrum cannot be fitted, grid searches can be used following the lines and the planes of the BZ. It should be notes that it is possible for 'symmetric' k vectors to occur within the degrees of freedom allowed for a given line or plane, by this we mean the presence of commensurate or related components rather the defined high symmetry points, and the restriction of possible planes may be appropriate if the components of the determined k vector from a grid search are close to a particular value or relationship.

The generation of the k vectors based on the BZ and subsequent conversion to the conventional unit cell chosen for the crystal structure also prevents mistakes. Firstly, the special nature of particular k vectors becomes apparent. Following this, the nature of an ordering wave vector, and its degrees of freedom become recognised as important physical parameters that distinguish a simple k vector based on a symmetry point from a locked-in k vector (based on an incommensurate line or plane in the BZ). The conversion to the conventional direct space cell, also prevents mistakes over the limits of possible k vectors that are frequent in centred cells, as the upper limit on the k vector along particular directions may be greater than unity.

### 3.2. Trialling of a given k vector

In order to reduce the susceptibility to false-positives that is intrinsic to conventional k vector search algorithms, the trialling of a given k vector is based on the physical scattering process.

The orientations of moments that are located at the crystallographic sites are refined against the experimental data using reverse-Monte Carlo (RMC) techniques. It is important to note that intensities are not extracted, but the full profile is used. This technique was first developed by us and released in SARA*h* [6] and prevents information loss due to extraction of intensities.

The number of RMC cycles that are required depends on the precise problem under investigation, though 50 -100 are typically enough to determine whether a k vector is able to fit the observed diffraction pattern.

### 3.3. Systems with several unrelated propagation vectors

One of the benefits of this methodology is that magnetic intensity will only be calculated at d-spacings suitable for moments at a given set of positions, with a particular k-vector. This creates a robust refinement strategy that can cope easily with difficult systems, such as those that contain several unrelated propagation vectors. Traditional methods, based on d-spacings alone do not discriminate the contributions of the different phases to the observed diffraction spectrum. In many cases the loss of information due to powder averaging and the use of only reflection positions makes the separation of different contributions untenable.

Using this more physical method, the presence of several unrelated propagation vectors becomes apparent and is signalled by the ability of a particular k vector to fit well a subset of the magnetic diffraction peaks and not others. An additional phase can then be introduced and used to determine the k vector of the next set of unfitted peaks. This procedure may be repeated as many times as required, and we have ourselves used this process to determine the propagation vectors of the magnetic field-induced ordering of the frustrated magnet GGG, where 2 commensurate and 1 incommensurate propagation vectors are simultaneously observed [8].

### 3.4. Systems with magnetic or non-magnetic impurity phases

Similarly, a dataset that has a small subset of peaks that cannot be refined may possess an impurity phase.[1] It should be remembered that with neutron scattering it is possible that magnetic impurity peaks may only be seen at low temperature, as the magnetic scattering may be stronger than the nuclear, depending on the details of the structure factors involved.

### 3.5. Brillouin zone indexing and k vector searches with SARA*h*

In the program SARA*h* the above procedure is implemented though a metaprogram structure with the Rietveld refinement program FullProf (FP) [12]. SARA*h* sets up the magnetic phase in the orientation matrix format based on the nuclear phase. The user may then automatically go through the different points, lines and planes of the BZ with SARAh automatically editing the FP input (pcr) file, launching the FP refinements, and performing RMC refinement of the moment orientations. The values of the k vector with the best fits, as characterised by the Rietveld $\chi^2$, are then listed. Particular k vectors chosen by the user may be automatically substituted into the FP input file and tested manually. Commensurate and incommensurate k vectors can be determined in this manner as SARA*h* automatically makes all relevant changes to the FP input file. Additional phases to describe additional k vectors can be added automatically as required.

---

[1] The observation of a small set of peaks suggests that untested propagation vectors, such as those missed in a general grid search of the Brillouin zone are not involved as these low symmetry k vectors typically have many reflections.

**Acknowledgements.** ASW would like to thank the Royal Society for funding.